\documentclass[twocolumn]{aastex701}

\usepackage[dvipsnames]{xcolor}

\newcommand{\vecx}{\mathbf{x}}
\newcommand{\solidwaffle}{\textsc{solid-waffle}}
\newcommand{\ipnlkernel}{$[K^2 a +KK']_{\Delta i, \Delta j}$}
\newcommand{\scanum}{20664}
\newcommand{\electron}{e$^-$}

\usepackage{duckuments}
\usepackage{lipsum}
\usepackage{comment}

\usepackage{amsmath}

\shorttitle{BFE in a Roman detector}
\shortauthors{Paine et al.}
\graphicspath{{./}{figures/}}

\begin{document}

\title{The Nancy Grace Roman Space Telescope Wide Field Instrument:\\Characterization of the Brighter-Fatter Effect in a Roman H4RG-10 Detector using the Talbot Effect}

\correspondingauthor{Jennie Paine}

\author[0000-0002-7517-9223]{Jennie Paine}
\affiliation{Center for Space Sciences and Technology, University of Maryland, Baltimore County, Baltimore, MD 21250, USA}
\affiliation{NASA Goddard Space Flight Center, Greenbelt, MD 20771, USA}
\affiliation{Center for Research and Exploration in Space Science and Technology, NASA/GSFC, Greenbelt, MD 20771, USA}
\email[show]{jennie.paine@umbc.edu}

\author[0000-0002-5636-233X]{Ami Choi}
\affiliation{NASA Goddard Space Flight Center, Greenbelt, MD 20771, USA}
\email{ami.choi@nasa.gov}

\author{Christopher Merchant}
\affiliation{NASA Goddard Space Flight Center, Greenbelt, MD 20771, USA}
\email{chrismerchant@hotmail.com}

\author[0000-0003-2630-8073]{Timothy D.~Brandt}
\affiliation{Space Telescope Science Institute, 3700 San Martin Drive, Baltimore, MD 21218, USA}
\email{tbrandt@stsci.edu}

\author[0000-0002-5861-7236]{Jeffrey Kruk}
\affiliation{NASA Goddard Space Flight Center, Greenbelt, MD 20771, USA}
\email{jeffrey.w.kruk@nasa.gov}

\author[0000-0001-5347-7062]{Joshua E. Schlieder}
\affiliation{NASA Goddard Space Flight Center, Greenbelt, MD 20771, USA}
\email{joshua.e.schlieder@nasa.gov}

\author[0000-0002-7567-4451]{ Edward~J.~Wollack}
\affiliation{NASA Goddard Space Flight Center, Greenbelt, MD 20771, USA}
\email{edward.j.wollack@nasa.gov}

\author[0009-0009-1006-604X]{Analia Cillis}
\affiliation{Center for Space Sciences and Technology, University of Maryland, Baltimore County, Baltimore, MD 21250, USA}
\affiliation{NASA Goddard Space Flight Center, Greenbelt, MD 20771, USA}
\affiliation{Center for Research and Exploration in Space Science and Technology, NASA/GSFC, Greenbelt, MD 20771, USA}
\email{analia.n.cillis@nasa.gov}

\author[0000-0002-5870-6108]{Jahmour J. Givans}
\affiliation{Department of Physics, University of Cincinnati, Cincinnati, OH 45221, USA}
\email{givansjj@ucmail.uc.edu}

\author{Robert Hill}
\affiliation{221 Hardwood Ct., Hardy, VA 24101}
\email{bob.hill87@gmail.com}

\author[0000-0002-2951-4932]{Christopher Hirata}
\affiliation{Center for Cosmology and Astroparticle Physics, The Ohio State University, 191 West Woodruff Ave, Columbus OH 43210, USA}
\affiliation{Department of Physics, The Ohio State University, 191 West Woodruff Ave, Columbus OH 43210, USA}
\affiliation{Department of Astronomy, The Ohio State University, 140 West 18th Avenue, Columbus OH 43210, USA}
\email{hirata.10@osu.edu}

\author[0000-0002-2457-272X]{Dana R. Louie}
\affiliation{Catholic University of America, Department of Physics, Washington, DC, 20064, USA}
\affiliation{NASA Goddard Space Flight Center, Greenbelt, MD 20771, USA}
\affiliation{Center for Research and Exploration in Space Science and Technology II, NASA/GSFC, Greenbelt, MD 20771, USA}
\email{dana.r.louie@nasa.gov}

\author[0000-0002-5982-566X]{Gregory Mosby}
\affiliation{NASA Goddard Space Flight Center, Greenbelt, MD 20771, USA}
\email{gregory.mosby@nasa.gov}

\author[0000-0002-2762-2024]{Anna Porredon}
\affiliation{CIEMAT, Avenida Complutense 40, E-28040 Madrid, Spain}
\email{annam.porredon@gmail.com}

\author[0000-0003-2662-6821]{Bernard J. Rauscher}
\affiliation{NASA Goddard Space Flight Center, Greenbelt, MD 20771, USA}
\affiliation{Rauscher Scientific LLC, Sykesville, MD 21784 USA}
\email{bernard@rauscherscientific.com}

\author[0000-0002-5073-3806]{Maxime J Rizzo}
\affiliation{NASA Goddard Space Flight Center, Greenbelt, MD 20771, USA}
\email{maxime.j.rizzo@nasa.gov}

\author[0009-0005-3569-9944]{Rujula Yete}
\affiliation{Department of Astronomy, University of Maryland, College Park, MD 20742, USA}
\email{rujuy5@gmail.com}

\begin{abstract}

The Nancy Grace Roman Space Telescope will conduct several community-driven surveys using its Wide Field Instrument (WFI), which contains 18 Teledyne H4RG-10 near-infrared detectors in the focal plane. Roman's wide area survey to measure weak gravitational lensing places particularly stringent requirements on point spread function (PSF) precision for the WFI. 
The brighter-fatter effect (BFE) is a nonlinear detector effect caused by the deflection of new charge away from pixels with previously accumulated charge during an exposure. As a result, the PSF measured at the detector depends on the charge contrast between pixels, and therefore the effective PSF for astronomical sources is a function of flux contrast and integration time.
We present characterization of the BFE observed in a non-flight H4RG by projecting a grid of bright point sources onto the detector using the Talbot Illuminator at the NASA Goddard Space Flight Center Detector Characterization Lab. We find that the effective area of a pixel decreases by $(-2.18\pm0.04) \times 10^{-6}$ per electron of charge contrast with its neighbors, corresponding to a measurable increase in the effective PSF size over the course of an exposure. Despite charge migration, the total signal detected in the vicinity of a bright source is roughly conserved. We also analyze flat field statistics for the same detector to measure the total inter-pixel nonlinearity (IPNL), which is dominated by BFE. We present a method to correct BFE in H4RG images using the flat field IPNL results, adapted from an extant optical correction technique. The correction reduces the observed BFE by about a factor of 25 as measured by the effective pixel area change, but there is remaining nonlinearity in the corrected data. We discuss possible sources of residual nonlinearity and suggest avenues to investigate optimal BFE correction in Roman survey images. 

\end{abstract}


\section{Introduction} \label{sec:intro}

The Nancy Grace Roman Space Telescope (Roman) is a NASA flagship mission designed to survey large areas of the sky in the near-infrared. One of Roman's primary science objectives is to study cosmic expansion and growth of structure via weak gravitational lensing, which entails measuring small (fraction of a percent) distortions in galaxy shapes. Achieving the science goals of weak lensing surveys requires thorough characterization and calibration of observational effects which bias galaxy shapes and subsequent cosmology. Thus weak lensing drives precision requirements for Roman's point spread function (PSF) calibration (PSF ellipticity known to $\leq 5.7\times10^{-4}$ RMS per component and size to $\leq 7.2\times10^{-4}$ relative RMS on angular multipole scales $\ell < 3200$; \citealt{SRD2023}), and motivates investigation into instrumental systematics which bias the PSF. 

Roman's High Latitude Wide Area Survey (HLWAS) is designed to conduct a weak lensing survey, utilizing the large field of view of the Wide Field Instrument (WFI) to measure the shapes of hundreds of millions of galaxies over a large area of the sky \citep{Schlieder2024,ROTAC2025}. The focal plane of the WFI contains 18 4k$\times$4k Teledyne H4RG-10 near-infrared detectors \citep{Mosby2020}. These detectors are capable of non-destructive readouts, enabling multiple reads of the detector signal in each pixel over the course of an exposure, referred to as sampling up the ramp. In an ideal detector, one would simply fit a slope to these ramps to determine the flux per pixel. However, there are multiple effects that cause a nonlinear relation between photons incident on the detectors and the measured signal. 

One detector nonlinearity of particular concern is the brighter-fatter effect (BFE; also sometimes referred to as charge migration), wherein the accumulation of charge on a pixel alters the electric field and deflects away new charge, causing the detected image to increase in size. Since the effect is dependent on integrated charge, it will be most apparent for bright sources over a given exposure time. The conventional model for BFE assumes that the effective collecting area of each pixel is dependent on charge contrast with its neighboring pixels \citep{Antilogus2014}: as charge accumulates over an exposure, the effective area of bright pixels decreases and the effective area of neighbors increases, resulting in charge migration away from bright pixels and nonlinear accumulation of charge. 
BFE was originally identified in optical charge-coupled devices \citep[CCDs; ][]{Antilogus2014,Guyonnet2015,Baumer2015,Gruen2015,Niemi2015}, and has subsequently been identified in near-infrared HXRGs \citep{PlazasMalagon2017,PlazasMalagon2018}. Previous studies have measured the effect in Roman H4RG detectors using flat field statistics \citep{Choi2020,Freudenburg2020,Givans2022} and point sources \citep{PlazasMalagon2024}. 
The results indicate that BFE, if left uncalibrated, will bias Roman's PSF beyond requirements for weak lensing; e.g.\ \citet{Choi2020} found that the BFE measured in an H4RG would modify the effective shape of a bright pixel by more than a percent in the J and H bands, exceeding the $7.2\times10^{-4}$ level calibration requirement. Thus, it is useful to further characterize the impact of BFE on H4RG measurements and to explore possible mitigation strategies.

In this paper, we present new analysis of point source images taken with a spare Roman detector to characterize the BFE. In Section~\ref{sec:Talbot Illuminator}, we describe the Talbot Illuminator, the testbed at NASA Goddard Space Flight Center which was used to produce a regular grid of spot images. In Section~\ref{sec:data}, we detail the spot images, flats, and darks used for our analysis. In Section~\ref{sec:IPNL}, we measure the total IPNL from flat field statistics, and in Section~\ref{sec:BFE measurements}, we characterize the BFE in spot images in terms of its effect on PSF size and nonlinearity around bright pixels. In Section~\ref{sec:correction}, we apply a BFE correction adapted from CCD-based correction methods to the spot data. Finally, we conclude and discuss future work to measure and correct BFE in Roman observations in Section~\ref{sec:discussion}.

\section{The Talbot Illuminator} \label{sec:Talbot Illuminator}

The Talbot Illuminator (TI) testbed was designed and built at NASA’s Goddard Space Flight Center in the Detector Characterization Laboratory. It utilizes the Talbot effect (also called self-imaging; \citealt{Talbot1836, Rayleigh1881}) to illuminate astronomical detectors with a 2-D array of point-like spots. 
The Talbot effect is a near-field diffraction phenomenon observed when light is transmitted through a periodic mask, causing the mask to be reimaged at half-integer multiples of the Talbot length $z_T=2a^2/\lambda$, where $a$ is the period of the mask and $\lambda$ is the wavelength of the incident light (see \citealt{Wen2013} for a review of the Talbot effect and its applications in optics). In the TI, a pinhole mask, described in more detail below, is used produce a rectilinear grid of spot images on the detector.
The system can generate spots as small as 1 $\mu$m in diameter, with tunable size and spacing, at wavelengths ranging from 480 nm to 2000 nm. The spot array can be translated laterally in two dimensions across the detector surface in sub-micron steps, permitting simultaneous measurement of the response of many pixels. 
This enables direct measurement of intra-pixel sensitivity variation, which was the original motivation for the design of the TI. However, the experimental design enables measurement of various detector characteristics of interest, including saturated point source response, persistence, and BFE.

\begin{figure}[t!]
\centering
\includegraphics[width=\columnwidth]{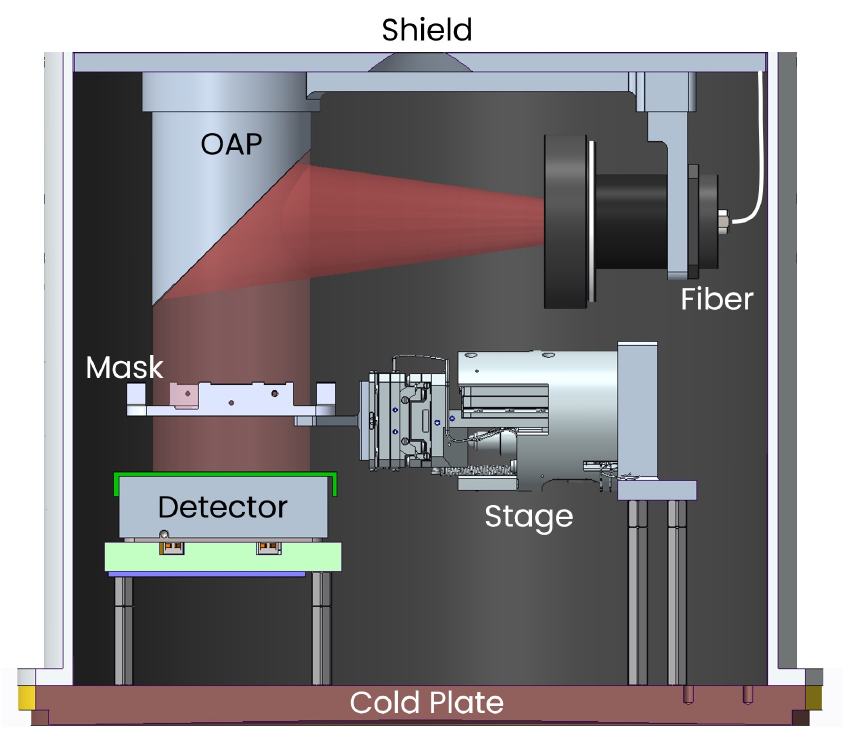}
\caption{A schematic of the Talbot Illuminator. Light from an external laser is fed into the cryostat by an optical fiber, collimated by the off-axis parabolic mirror (OAP), and then passed through a photomask. The mask contains regularly spaced pinholes such that the interference pattern produced at the detector is a rectilinear grid of spots described by the Talbot effect. The cryogenic translation stage is used to adjust the distance between the mask and the detector, and can be used to translate the detector laterally.\label{fig:TI schematic}}
\end{figure}

A schematic of the test setup is given in Figure~\ref{fig:TI schematic}. 
Light from an external laser is fed into the cryostat via an optical fiber, and all other components necessary for generating spots on the surface of a detector are contained within the cryostat. The laser light is collimated by the off-axis parabolic mirror (OAP) and then passed through a pinhole mask.
An interference pattern which can be described by the Talbot effect is created at the detector surface. 
Mask-to-detector distance is adjusted using a cryogenic translation stage. When the distance is equivalent to the Talbot length for the given wavelength and aperture spacing, the interference pattern produces the reimaged periodic spot mask with minimized spot size on the detector.

\begin{figure}[t!]
    \centering
    \includegraphics[width=\columnwidth]{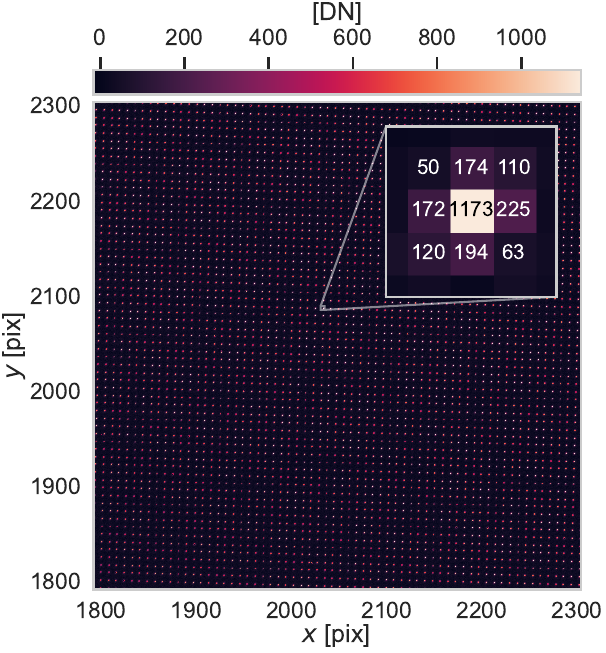}
\caption{Example spots from the Talbot Illuminator testbed, taken from the CDS image $S_{10} - S_{9}$ of the first exposure of the saturation test sequence detailed in Sec.~\ref{sec:talbot data}. The inset shows a spot that is nearly centered on a pixel and the pixel values in DN for the nine pixels centered on the spot. \label{fig:spots_example}}
\end{figure}

In this work, we utilize data taken with a 480 nm continuous wave laser source, and a photomask with 2 $\mu$m pinholes and 80 $\mu$m spacing. The mask was produced by Front Range Photomask\footnote{\url{https://www.frontrangephotomask.com}} with a 150 nm Cr layer over a glass substrate.
Figure~\ref{fig:spots_example} shows an example correlated double sample (CDS) image of the spot array generated by the TI in this set up. The detector signal is given in units of data numbers (DN), and we use $S_i$ to denote the signal in the $i$th readout, where $i=0$ is the first frame after the reset read. 
Spots are created with 80 $\mu$m spacing in both x and y directions, which is a result of the mask design. This means there are approximately 15,000 spots generated on a 1000$\times$1000 pixel array. A Moiré pattern is created because of the misalignment between the mask and detector pixel locations. The spot-background ratio varies depending on the mask used in the TI setup and the wavelength of light. For this mask and the 480 nm laser, the resulting ratio between the central pixel value and its nearest neighbor is approximately 6:1, and the ratio between the central pixel value and the background value is approximately 45:1. 

\section{Data} \label{sec:data}

All images utilized in this work were taken with detector SCA 20664, an engineering grade device that was screened as part of the Roman WFI detector test program but was not prioritized for flight or as a flight spare.
We primarily focus on spot images produced by the TI, and use flats and darks taken in the TI testbed for initial processing of the spot images, which are described below in \ref{sec:talbot data}. However, for the purpose of measuring a flats-based IPNL kernel, we utilize a separate sequence of flat and dark images, described in \ref{sec:flats and darks}.

\subsection{Spot images from the Talbot Illuminator} \label{sec:talbot data}

\begin{figure}[t!]
    \centering
    \includegraphics[width=\columnwidth, height=\columnwidth]{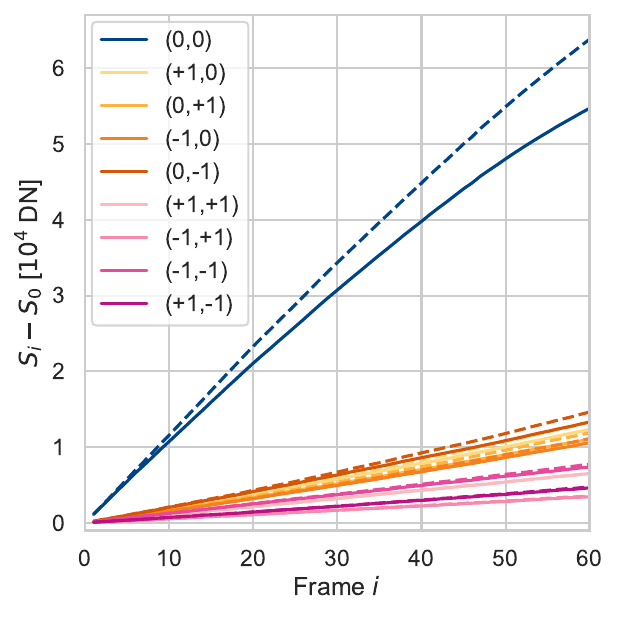}
\caption{Ramps before (solid lines) and after (dashed lines) CNL correction for the example spot featured in the Figure \ref{fig:spots_example} cutout. Blue represents the brightest pixel of the spot (0,0), shades of orange represent the 4 nearest pixels, and shades of pink represent the 4 diagonal pixels.  \label{fig:ramps_example}}
\end{figure}

The TI data utilized in this paper were part of a sequence of darks and spot images intended to test the saturation response of the detector. The TI testbed was set up with a 480 nm laser, shone through a 2 $\mu$m pinhole mask to produce spot images with 80 $\mu$m spacing. The detector, SCA 20664, was connected to a flight-like ACADIA controller \citep{Loose2018}. Each exposure in the test had 201 frames and 2.99 s frame time. The test began with 5 dark exposures, followed by a repeating sequence of two spot exposures and two dark exposures, with increasing spot flux after each set of darks. Flux was controlled by adjusting the laser attenuation, though we note that the TI does not produce uniform spot fluxes across the detector. For this work, we use only the second set of two spot exposures in the sequence, where the target flux per spot was about 370 \electron/s, resulting in spots which approach saturation around frame 60. This gives us about 50 frames per exposure to assess the BFE prior to the onset of hard saturation. An example spot image from these exposures is shown in Figure \ref{fig:spots_example}, and ramps for pixels around one of the spots is shown in Figure \ref{fig:ramps_example}.

\subsubsection{Classical nonlinearity correction} \label{sec:cnl}

Roman's H4RG detectors exhibit classical nonlinearity (CNL), such that the accumulated counts in a pixel are nonlinear in time when exposed to a constant flux of photons \citep{Mosby2020}. CNL is a collective term for several nonlinear effects, but in principle does not include inter-pixel nonlinearities such as BFE. A common technique to correct CNL is to assume that the measured counts are an $N$th order polynomial function of the true counts, and to derive the polynomial coefficients per pixel by fitting to a sample of ramps. We employ the specific CNL correction algorithm of \citet{Brandt2025} \footnote{\url{https://github.com/RomanSpaceTelescope/SOCReferenceFileCode}}.

We fit CNL coefficients using two exposures of diffuse illumination on SCA \scanum\ taken in the TI testbed prior to the spot saturation sequence. For these exposures, a second fiber was positioned inside the cryostat so that it would illuminate around the photomask, as opposed to directly behind the mask as for the typical spot exposures. The light source was a 1310 nm laser operated at low power. The image resulting from the diffuse illumination is not exactly flat due to the off-center position of the light source and reflection off of other components of the TI setup, but the images do not need to be completely uniform to derive CNL coefficients. Both exposures were 501 frames and reached saturation around frame 300, resulting in a large sample of reads per pixel from which to derive the CNL correction. 
We fit a 4th order polynomial correction to each pixel using only counts up to $6\times10^4$ DN, which is below digital saturation. We applied the CNL correction to the spots and dark ramps, and then subtracted the median dark ramp from the spot ramps. Figure \ref{fig:ramps_example} shows ramps before and after applying this correction procedure for pixels around an example spot. The CNL-corrected ramps are still not strictly linear, and in the analysis below we assume that the remaining nonlinearity after CNL correction can be mostly attributed to the BFE. Details of how the CNL correction impacts the subsequent BFE results are in Appendix~\ref{app:cnl}.

\subsection{Additional flats and darks} \label{sec:flats and darks}

We also use flats and darks taken with the same detector during acceptance testing in the Detector Characterization Laboratory to measure the IPNL kernel across the detector using the flat field autocorrelation approach. The data were taken in the same fashion as those described in Section 5.1 of \cite{Freudenburg2020}, but the data for detector 20664 were not utilized in their work. To summarize, the detector was operated with 32 output channels, 1.0 V bias, and guide windows turned off for a sequence of darks and flats. Unlike the TI data which were taken with an ACADIA controller, these data taken during acceptance testing used a Leach controller. The sequence includes sets of variable numbers of exposures, alternating between sets of darks and sets of flats (see Figure 3 from \citealt{Freudenburg2020}). We utilized exposures from the first four sets in the test sequence, which consisted of 10 darks, 5 flats, 8 darks, and 11 flats, respectively. Each exposure had 64 frames and 2.764 s frame time. The flats had fluxes per pixel of about 1100 \electron/s, and saturate around halfway through each exposure.

\section{Inter-pixel nonlinearity measurement from flats} \label{sec:IPNL}

\begin{figure*}[t!]
    \centering
    \includegraphics[width=\textwidth]{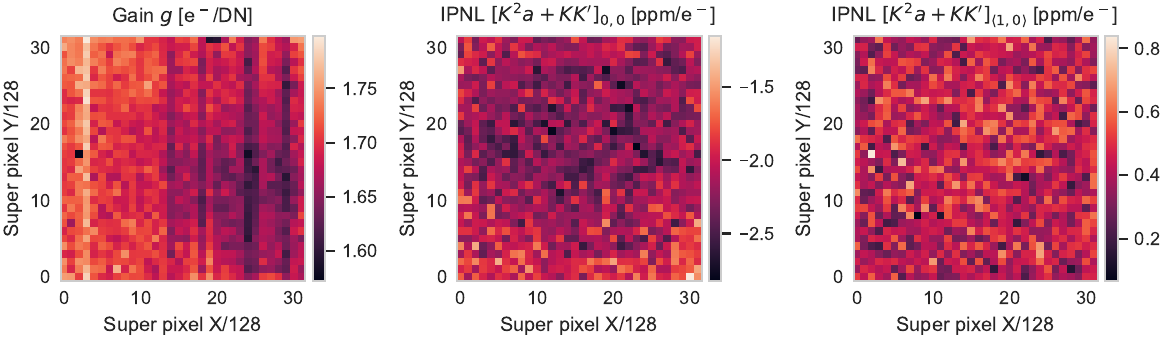}
    \caption{Maps of the gain and IPNL for SCA \scanum\ produced by flat field statistics analysis. \label{fig:solid-waffle maps}}
\end{figure*}

\begin{deluxetable}{ccc}
\tablewidth{0pt}
\tablecaption{IPNL results \label{tab:ipnl}}
\tablehead{
\colhead{\ipnlkernel} & \colhead{Mean} & \colhead{Standard deviation} \\
\colhead{} & \colhead{(ppm/e$^-$)} & \colhead{(ppm/e$^-$)} }
\startdata
$0,0$ & -2.142 & 0.232 \\
$\langle 1,0 \rangle$ & 0.437 & 0.107 \\
$\langle 1,1 \rangle$ & 0.113 & 0.103 \\
$\langle 2,0 \rangle$ & 0.033 & 0.105 \\
$\langle 2,1 \rangle$ & 0.014 & 0.074 \\
$\langle 2,2 \rangle$ & 0.004 & 0.108 \\
\enddata
\tablecomments{IPNL coefficients for SCA \scanum, averaged over all super-pixels. $\langle \Delta i , \Delta j \rangle$ are the mean over all pixels at the same separation from the central pixel; e.g.\ $\langle 1 , 0 \rangle$ are the nearest neighbors, $\langle 1 , 1 \rangle$ are diagonals, etc.}
\end{deluxetable}

Inter-pixel nonlinear effects like BFE manifest in flat field images as a deviation from purely Poisson noise due to inter-pixel correlations. To measure this, we utilize \solidwaffle\footnote{\url{https://github.com/Roman-HLIS-Cosmology-PIT/solid-waffle}}, a code repository for correlation analysis in HXRG flats, which was described and tested on Roman detectors in a series of papers. The formalism for measuring BFE from H4RG flat field correlations was described by \citet{Hirata2020}, and the code was first applied to a non-flight prototype Roman detector by \citet{Choi2020}. \citet{Freudenburg2020} introduced a Fourier-space treatment of the correlations, and \citet{Givans2022} further expanded the formalism to include quantum yield and charge diffusion. This formalism was applied to three flight lot detectors, including two that were ultimately installed in the focal plane \citep{Schlieder2024}. We direct the reader to those papers for full descriptions of the theoretical background and detector characterization code, but briefly summarize the parameters of interest to our analysis. 

The BFE model \citep{Antilogus2014} assumes that a pixel located at column and row ($i,j$) has an effective collecting area that is dependent on charge as
\begin{equation}
    \mathcal{A}_{ij} = 
    \mathcal{A}^0_{ij} \left[ 1 + \sum_{\Delta i, \Delta j} a_{\Delta i, \Delta j} Q(i+\Delta i , j+ \Delta j) \right],
\end{equation}
where $\mathcal{A}^0_{ij}$ is the original pixel area, $a_{\Delta i, \Delta j}$ is a matrix representing the coupling between pixels, and $Q(i,j)$ is the charge in the pixel. The BFE matrix $a_{\Delta i, \Delta j}$ is often expressed in units of ppm/e$^-$. 

Another significant source of coupling between pixels is inter-pixel capacitance (IPC), in which the voltage on a pixel is sensitive to the charge in neighboring pixels. The effect is modeled by a cross-talk kernel $K$, which has both linear and nonlinear components and is normalized such that $\sum_{\Delta i, \Delta j} K_{\Delta i, \Delta j} = 1$. 
The signal level-dependent nonlinear IPC kernel is parameterized by $K'$.
Note that IPC and BFE occur at different stages of the detection process: BFE affects the collection of charge at the pixels, whereas IPC affects the conversion of collected charge to voltage.
The combined inter-pixel effects of linear IPC, nonlinear IPC, and BFE, are represented by a $5\times5$ IPNL kernel, \ipnlkernel.
Previous work has shown that the measured IPNL is dominated by BFE in Roman's H4RG detectors \citep{Choi2020, Freudenburg2020}. 

The autocorrelation analysis pipeline characterizes various detector parameters averaged over super-pixels, including gain, CNL, IPC, and BFE using a series of flats and darks. 
For consistency with previous works, we followed the ``fiducial" analysis of \citet{Freudenburg2020} on the sequence of flats and darks described in Section~\ref{sec:flats and darks}. Results are averaged over 16 flats taken in the second and fourth sets of exposures, and paired with 16 darks taken in the first and third set of exposures. The arrays are binned into $32\times32$ super-pixels ($128\times128$ pixels per super-pixel). IPNL is determined using the non-overlapping correlation function from frames 1, 5, 6, and 10. We configured \solidwaffle\ to perform advanced characterization over 3 iterations, which iteratively corrects IPNL bias to other detector parameters, including gain, IPC, and classical nonlinearity. 

The mean values and standard deviations over all super-pixels of the resulting IPNL parameters are listed in Table~\ref{tab:ipnl}. We also show the central IPNL, nearest neighbor IPNL, and gain maps in Figure~\ref{fig:solid-waffle maps}. 
The central pixel area decreases as a function of collected charge by  $-2.142 \pm 0.007$ ppm/e$^-$. The nearest neighbor pixel area increases by $0.437 \pm 0.003$ ppm/e$^-$, which \citet{Freudenburg2020} found to be a more robust IPNL parameter than the central pixel value. 
Though SCA \scanum\ is not integrated into the WFI and we expect some variation in IPNL between detectors, these results are comparable to those found for flight detectors SCA 20663 and SCA 20829, which are labeled WFI06 and WFI12 in the focal plane array. The mean and standard deviation gain over all super-pixels were $1.6232\pm0.0469$ and $1.7285\pm0.0450$ \electron/DN for SCA 20663 and SCA 20829, respectively; the mean and standard deviation of the nearest neighbor IPNL were $0.4396\pm0.1384$ and $0.3537\pm0.0930$ ppm/\electron\ \citep{Freudenburg2020}.

\section{BFE measurements from spot images} \label{sec:BFE measurements}

In this section, we implement two complementary methods to measure and characterize the impact of BFE in SCA 20664 utilizing the CNL-corrected spot images from the Talbot Illuminator. First is a demonstration of the increasing size of the point spread function as measured from the spots due to charge accumulation (\ref{sec:epsf}), and second is a measurement of the additional nonlinearity in pixels containing and neighboring spots (\ref{sec:flux deviation}).

\subsection{Growth of the effective PSF} \label{sec:epsf}

\begin{figure*}[ht!]
\centering
\includegraphics[width=0.8\textwidth]{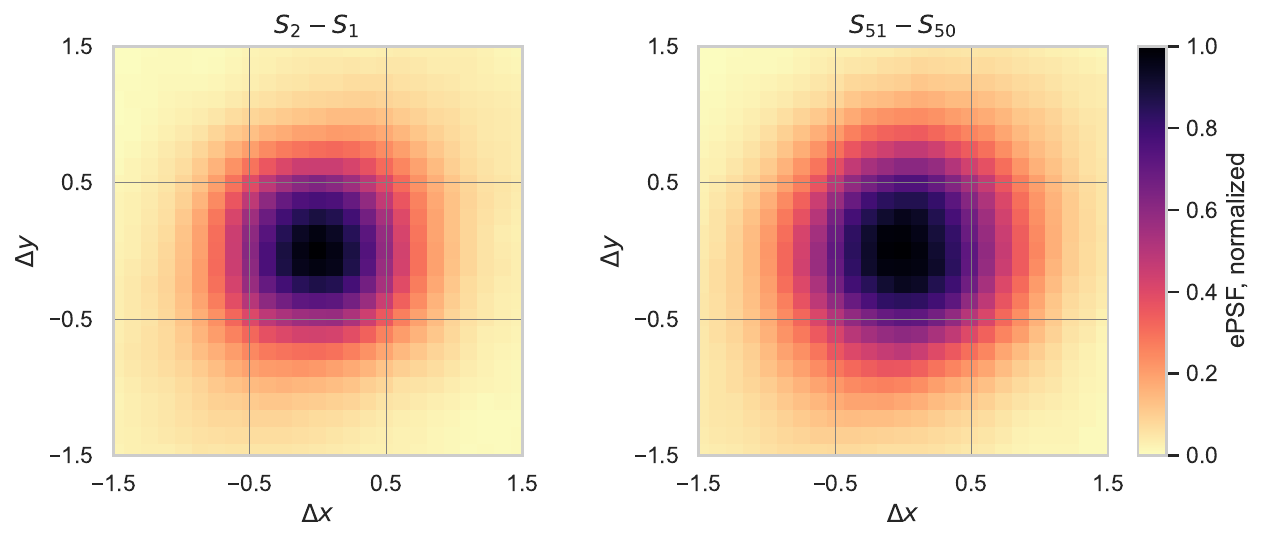}
\caption{ Example ePSF models from the first (left) and last (right) single frame difference images used in the analysis. Both models were produced from the first spot exposure in the sequence and using the 204 spot sample located between pixel coordinates 2048-2176 in x and y. The images are cropped to the $3\times3$ pixel region around the center of ePSFs, and normalized so that the maximum value in each image is 1.  \label{fig:ePSF}}
\end{figure*}

While the optical PSF of a telescope should not depend on flux, the effective PSF (ePSF) measured from the detectors may vary with source flux and pixel signals due to nonlinearities such as the BFE. The ePSF, as defined by \cite{Anderson2000}, is the convolution of the optical PSF and the pixel response function; i.e. it is a model of the fraction of a point source's light that will be detected in a particular pixel. In principle, one would expect the BFE to result in wider ePSF models for higher flux sources, and that the ePSF will increase in size over the course of a single exposure as charge accumulates in the pixels. Here we investigate this expectation using spot images from the Talbot Illuminator.

Spots were initially detected in each exposure using the $S_{5}-S_1$ CDS image, and their initial centroids determined by fitting a 2D Gaussian to each spot. 
The TI images show some spatial variation in spot brightness, with the brightest spots generally near the center of the detector, and the large scale brightness variations are consistent between exposures. Since we are trying to constrain an effect which depends on both flux and integrated signal, we divide spots into $128\times128$ pixel regions and limit our ePSF analysis to only the 16 regions within $\pm 256$ pixels from the center of the detector (i.e. between pixel coordinates 1792 and 2304 in both x and y). We initially estimate spot flux from the total signal in a 3 pixel radius circular aperture around each spot in the $S_{5}-S_1$ image after subtracting the median background signal. These frames were selected to measure the accumulated signal over several frames early in the exposure, when the effect of BFE should be minimal. The 3 pixel radius aperture was chosen by inspecting the aperture sums resulting from radii between 1-4 pixels in 0.5 pixel width increments. The median total signal is stable for radii $\geq3$ pixels, and choosing a radius $<4$ pixels ensures that the apertures do not overlap for spots with $\sim$8 pixel separation. In each region, we implement an iterative 3$\sigma$ clip on the initial aperture photometry to exclude outlying bright or faint spots. The resulting samples include between 198-216 spots per $128\times128$ pixel region.

We next fit ePSF models to the 16 groups of spots and each subsequent frame pair difference image, $S_{i+1}-S_i$ from frame $i=1$ to 50. ePSF models are fit using the \textsc{photutils} python package \citep{Bradley2025} and its \textsc{EPSFBuilder} class, which follows the methodology detailed in \cite{Anderson2000} and \cite{Anderson2016}. We select $7\times7$ pixel cutouts around each spot and perform local background subtraction on the cutouts. The ePSF models are then fit using an oversampling factor of 8. 

The ePSF fitting process is iterative, wherein each new iteration takes the centroid positions and fluxes of the spots from the previous iteration as input. We also align the spot centroids to a rotated grid between each iteration: after spots are fit with the latest ePSF model using the \textsc{EPSFFitter} class, we match the spots to a rectilinear grid and perform a least-squares affine fit to transform the unit grid to the measured spot positions. This process uses the grid-solution to improve the accuracy of each spot centroid in fewer iterations than measuring each spot centroid independently, and we find that a maximum of five iterations is sufficient. The results are an over-sampled ePSF model per frame in each subregion of the detector, and an ePSF-fit position and flux per frame for each input spot.

\begin{figure*}[t!]
\centering
\includegraphics[width=\textwidth]{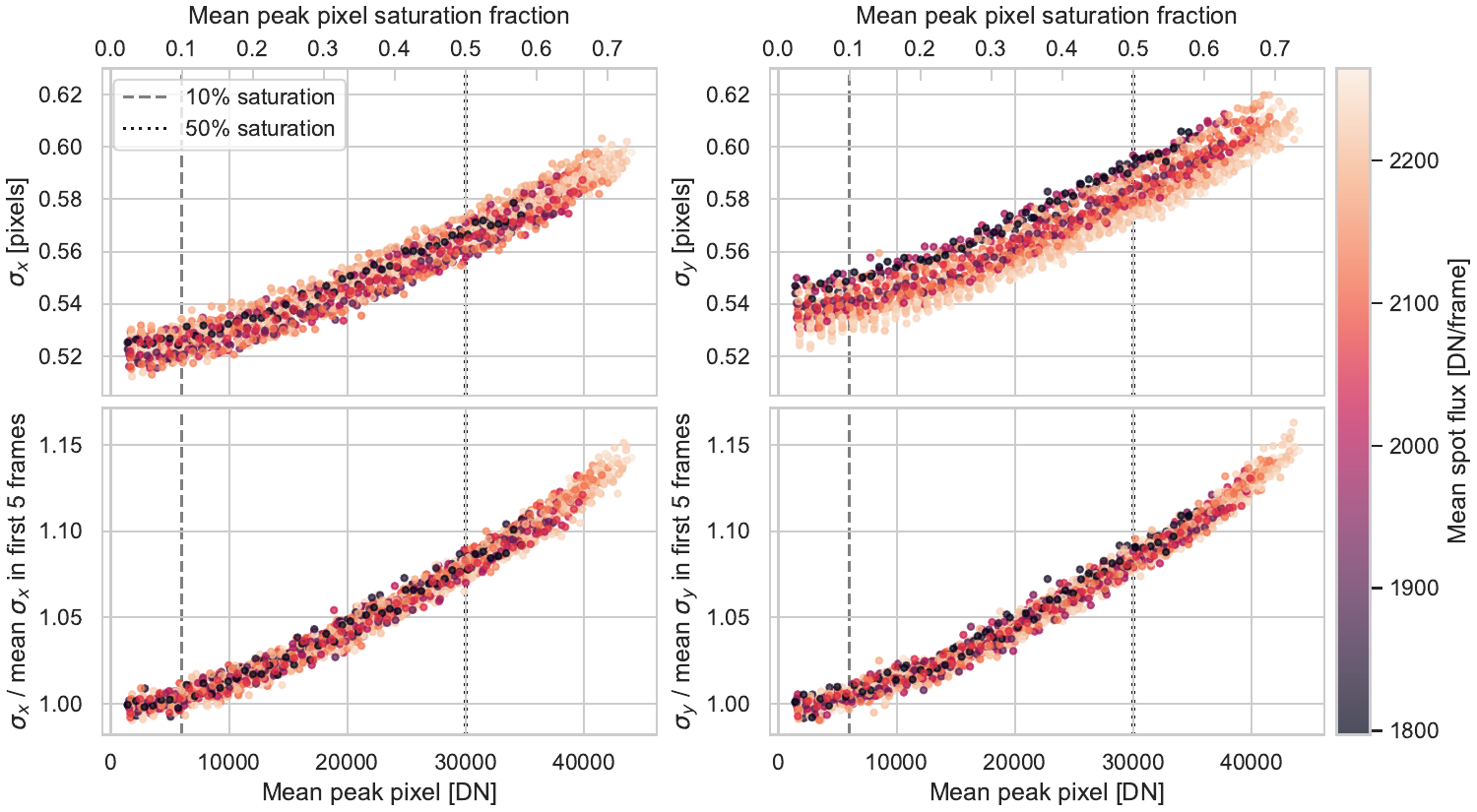}
\caption{Top row: standard deviation of 2D Gaussian fits to the ePSFs in the detector $x$ and $y$ directions vs.\ the mean maximum pixel value of spots sampled to fit the ePSF. Bottom row: the same data as the top row, with the sizes scaled by the mean size in the first 5 frames.  The color bar indicates the mean spot flux in each subregion of the detector. The absolute size of the ePSF varies by subregion and is asymmetric in $x$ and $y$, but the size relative to the initial several frames (which is assumed to be closest to the ideal ePSF) appears to grow at about the same rate as signal increases. \label{fig:ePSF sizes and growth}}
\end{figure*}

In Figure~\ref{fig:ePSF}, we show example ePSF models for the subset of spots located between pixel coordinates 2048-2176, which contained 204 spots with a typical flux of $\sim$2200 DN/frame. The example ePSFs were fit to the $S_2-S_1$ and $S_{51}-S_{50}$ CDS images of the first spots exposure in the TI test sequence. 
The ePSFs show a characteristic increase in size between the early and late frames due to BFE. 
To quantify the change in ePSF size as signal increases, we fit a 2D Gaussian model to each ePSF. We note that the true spot profiles are not exactly Gaussian, but the Gaussian width provides an estimate of the size of the ePSFs. 
Figure \ref{fig:ePSF sizes and growth} shows the standard deviations, $\sigma_x$ and $\sigma_y$, in the detector x and y directions as a function of the mean peak pixel value of the input spot data. The spots are slightly larger in the detector y direction than x (the FWHM in early frames is about 1.22 pix in x and 1.27 pix in y). There is also an apparent correlation between $\sigma_y$ and spot flux, with brighter spots having smaller $\sigma_y$ than fainter spots, which may be a result of the spot mask and unrelated to detector effects. 

Figure \ref{fig:ePSF sizes and growth} also shows the relative change in ePSF size as a function of signal. Each subset of spot sizes appear to follow the same trend as a function of integrated signal regardless of the typical spot flux or differences in initial size. Notably, this broadening is apparent early in the ramp, and by about 50\% of full-well the ePSF has grown by $>6$\%. The ePSF growth is apparently slower at lower signals, indicating that the BFE is not strictly linear with signal. We explore this further in Section~\ref{sec:correction} when we test a linear BFE correction on the data.

\subsection{Deviation from constant flux around point sources} \label{sec:flux deviation}

\begin{figure*}[t!]
    \centering
    \includegraphics[width=\textwidth]{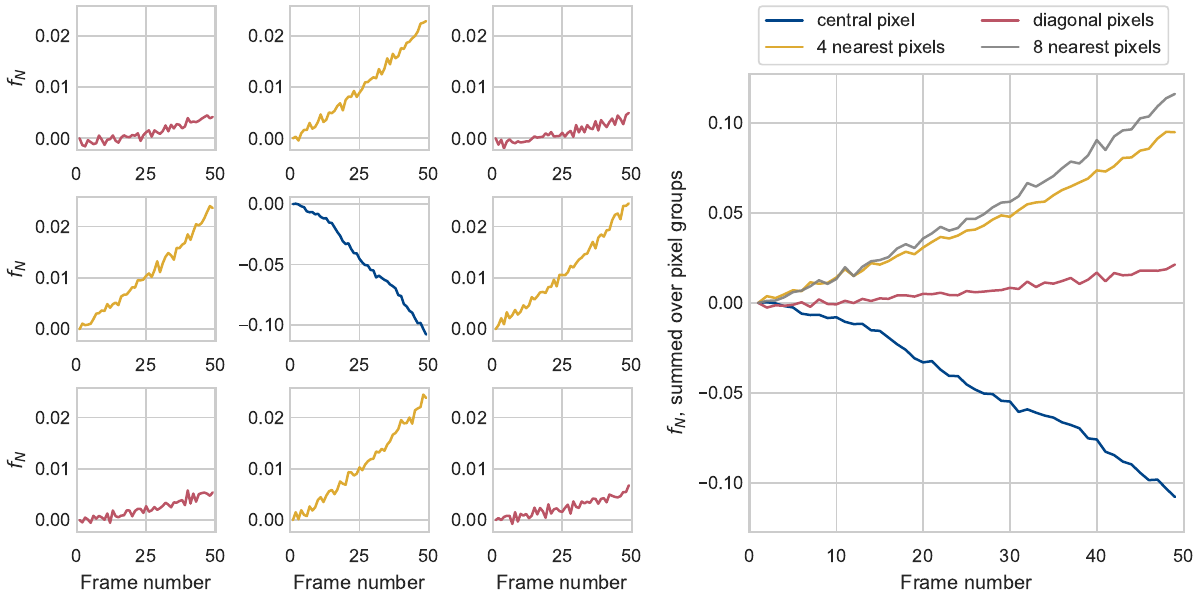}
    \caption{Mean relative flux deviation, $f_N$, in $3\times3$ pixel regions around spots with centroids within 0.1 pixel widths from a pixel center. The left plots show the median flux deviation per pixel in the $3\times3$ pixel grid. Note that the vertical axis scale is different for the central pixel compared to the eight neighboring pixels in the left figure. The right plot shows the same data summed over pixel groupings with respect to distance from the central pixel: dark blue for the central pixel, orange for the four nearest neighboring pixels, pink for diagonal pixels, and gray for the total of the eight neighboring pixels.  \label{fig:flux deviation}}
\end{figure*}

We now consider how BFE changes the flux in individual pixels due to charge contrast in neighboring pixels. 
BFE will cause the rate of charge accumulation in bright pixels to decrease as total charge increases, and this should be conserved by increasing the rate of charge accumulation in neighboring pixels. To measure this changing flux in our spot data, we adopt the methodology of \cite{PlazasMalagon2024}, who define the relative flux deviation in a pixel as 
\begin{equation}
    f_N(i) = \frac{F_i - F_1}{F_*}, 
\end{equation}
where $F_i$ is the flux in a pixel calculated from the difference in signal per frame time $\Delta t$ between frames $i+1$ and $i$, i.e.\ $F_i(x,y) = [S_{i+1}(x,y) - S_{i}(x,y)]/\Delta t$.
$F_*$ is a normalization factor defined for $3\times3$ pixel cutouts around each spot as 
\begin{equation}
    F_* = \frac{\sum_{x,y} \left[S_{M-1}(x, y) - S_{1}(x, y)\right]}{(M-2)\Delta t}  , 
\end{equation}
where $M$ is the total number of frames. In an ideal detector, $f_N$ should be zero, but \cite{PlazasMalagon2024} showed that the BFE causes $f_N$ to trend negative in bright pixels and positive in neighboring, lower signal pixels. 

We calculated the relative flux deviation per pixel using a sample of spots in the same central region of the detector used for the ePSF analysis in Section~\ref{sec:epsf} (i.e.\ within $\pm256$ pixels from the center of the detector). We selected only spots where the ePSF-based position is centered on a pixel to within 0.1 pixel widths. Pixel-centered spots were selected because we want to average the results over many spot cutouts, but spots located closer to a pixel boundary would be expected to result in an asymmetric change in measured flux in neighboring pixels depending on the direction of higher contrast. We also excluded spots neighboring a masked pixel. This resulted in a sample 98 spots, and for each spot we measured the flux deviation in the two exposures separately. 

To measure the relative flux deviation for each pixel-centered spot, we extracted a $3\times3$ pixel cutout from the CNL-corrected and background subtracted image cubes, and calculated $f_N$ per pixel up to the 50th frame. The mean $f_N$ in each pixel, averaged over spots in both exposures, are shown in Figure \ref{fig:flux deviation}. The figure also  demonstrates that flux lost in the central brightest pixel is roughly conserved via increasing flux in the 8 surrounding pixels due to charge migration, with most of the increasing flux in the 4 nearest neighboring pixels. 

Assuming the effective pixel area change model for BFE, \cite{PlazasMalagon2024} defined the parameter $B$ as the fractional pixel area change per electron of charge contrast with neighboring pixels. It is related to the slope of $f_N(i)$ as 
\begin{equation}
    f_N(i) = \frac{BF_1F_c\Delta t}{F_*} i,    
\end{equation}
where $F_c$ is the mean flux contrast between the center pixel and its nearest neighbors. This definition assumes that the incident photon flux is uniform across the central pixel, which is not true for our spot images. We also note that the value of $B$ is correlated with gain. We used the gain maps produced with \textsc{solid-waffle} in Section~\ref{sec:IPNL} to convert from DN to electrons for consistency in comparing $B$ to the flat field IPNL kernel results. 

We fit a line to $f_N(i)$ in the central pixel for each spot, and calculated $B$ from the resulting best-fit slope. The mean value of $B$ is $-2.18\pm0.04$ ppm/e$^-$, and the standard deviation of the distribution is 0.61 ppm/e$^-$. Note that the standard error on $B$ does not account for uncertainty due to the shape of the PSF and the assumption of uniform illumination in the central pixel.

\section{Application of IPNL to correct spot images} \label{sec:correction}

As demonstrated above, BFE causes significant nonlinearity and changes to the effective PSF measured by a Roman detector over the course of an exposure, and these effects will translate to bias in Roman science which is sensitive to PSF calibration if the BFE is not calibrated or corrected. Weak lensing measurements are particularly sensitive to PSF calibration, and specifically require using the PSF calibrated on bright, high contrast stars to measure shapes of fainter, lower contrast galaxies. For this reason, there have been previous efforts to correct BFE for weak lensing experiments using CCDs \citep{Coulton2018,Astier2023,Broughton2024}. These BFE correction algorithms generally rely on translating flat field statistics to a correction for science data using the change in pixel area due to BFE; for example, \citet{Coulton2018} model BFE using a scalar kernel that can be effectively deconvolved from an image, and \citet{Astier2023} apply an electrostatic model derived from flats-based area change coefficients. 

\subsection{Adapted scalar kernel correction}

We present a BFE correction adapted from \citet{Coulton2018}, using the flats-based IPNL matrix (Section~\ref{sec:IPNL}) to correct BFE in the spot images. \citet{Coulton2018} proposed that distortions due to BFE may be corrected in CCD images by approximately deconvolving the BFE kernel measured from flat field covariances. Under the assumption that the incident intensity of charge $I(\vecx)$ is subject to a deflection field due to a constant, translationally invariant BFE kernel $K$, the measured image after some total integration time $t$ is  
\begin{equation}
    \begin{split}
    \langle Q(\vecx,  & t) \rangle  = tI(\vecx) - \frac{t^2}{2} \frac{\partial}{\partial x_i} \\
    & \times \left(  I(\vecx) \frac{\partial}{\partial x_i} \int d^2\vecx' K(\mathbf{x-x'})I(\mathbf{x'}) \right).
    \end{split}
\end{equation} 
The true image can then be estimated from the measured image by 
\begin{equation}
\begin{split}
\hat{Q}(\vecx&, t) = Q(\vecx, t) \\
& + \frac{1}{2} \left(  Q(\vecx, t) \frac{\partial}{\partial x_i} \int d^2\vecx' K(\mathbf{x-x'})Q(\mathbf{x'},t) \right)
\end{split}
\end{equation} 
(see Eq. 22 of \citealt{Coulton2018}). The correction term is the average effect over the integration time, assuming that the first charge experiences no deflection due to BFE and that the last charge experiences twice the average effect. 
Note that the deflection field is dependent on integrated charge, but that the strength of the BFE, parameterized by the kernel $K$, is not a function of charge. This assumption of constant BFE is not necessarily true. In fact, \citet{Broughton2024} showed that the BFE in LSST's CCDs is not linear in flux and the scalar kernel correction does not account for this higher-order BFE. 

For Roman's H4RG detectors, we may use the samples up the ramp to correct the signal in each frame as opposed to total integrated signal, and therefore reconstruct the corrected ramps while relaxing assumptions about the time evolution of the detected image. From Eq.~12 of \citet{Coulton2018}, we can relate the collected charge after a small increment of time $dt$ to the incident charge and the deflection field due to previously accumulated charge:
\begin{equation}
    \begin{split}
    \langle Q(\vecx,  & t+dt) \rangle  = \langle Q(\vecx, t) \rangle + I(\vecx) dt \\
    & - \frac{\partial}{\partial x_i} \left(  I(\vecx) \frac{\partial}{\partial x_i} \int d^2\vecx' K(\mathbf{x-x'})\langle Q(\mathbf{x'},t) \rangle \right)dt.
    \end{split}
\end{equation} 
The first two terms are the expected change in charge for an ideal detector, and the third term describes the deflection of new charge due to the convolution of the BFE kernel with the accumulated charge in the pixels. 
From this expression, we can write down the measured flux over a single frame, i.e.~$F_i(\vecx) = [Q_{i+1}(\vecx) - Q_{i}(\vecx)] / \Delta t$, as
\begin{equation}
    \begin{split}
    \langle F_i(\vecx) \rangle &= I(\vecx) \\
    & - \frac{\partial}{\partial x_i} \left(  I(\vecx) \frac{\partial}{\partial x_i} \int d^2\vecx' K(\mathbf{x-x'})\langle Q_i(\mathbf{x'}) \rangle \right).
    \end{split}
\end{equation}
One may then estimate the true flux per frame as
\begin{equation}
    \begin{split}
    \hat{F}_i(\vecx) &= F_i(\vecx) \\
    &+ \frac{\partial}{\partial x_i} \left(  F_i(\vecx) \frac{\partial}{\partial x_i} \int d^2\vecx' K(\mathbf{x-x'})Q_i(\mathbf{x'}) \right).
    \end{split}
\end{equation}

As in \citet{Coulton2018}, we apply the correction iteratively: we calculate the correction term $\nabla [ F_i(\vecx) \nabla (K \ast Q_i(\vecx)) ] $, add it to the measured flux $F_i(\vecx)$, and repeat until the total added charge in an iteration falls below 10 e$^-$. Note that we do not modify the accumulated charge image $Q_i(\vecx)$ in each step. The charge deflection at any time is dependent on the incident flux of charge, which we aim to reconstruct, and the actual charge accumulated at the pixels that has already experienced the effect of the BFE due to the preceding charge. Therefore, in each iteration we change only the measured flux per frame in order to converge on an estimate of the true incident flux. The corrected ramp is then simply the cumulative sum of the corrected flux in e$^-$ per frame. 

We applied this correction to the CNL-corrected spot data after converting from DN to e$^-$ using the gain map fit by \textsc{solid-waffle} (Section~\ref{sec:IPNL}). Since BFE apparently dominates the measured total IPNL \citep{Choi2020, Freudenburg2020}, we interpret the IPNL matrix, \ipnlkernel, as the BFE kernel, and apply the mean IPNL matrix over all super-pixels (Table~\ref{tab:ipnl}) in the BFE correction.
We then conducted the same analysis described in Section~\ref{sec:BFE measurements} using the BFE-corrected ramps to assess the quality of the correction.

\subsection{Results}

\begin{figure}[t!]
\centering
\includegraphics[width=\columnwidth]{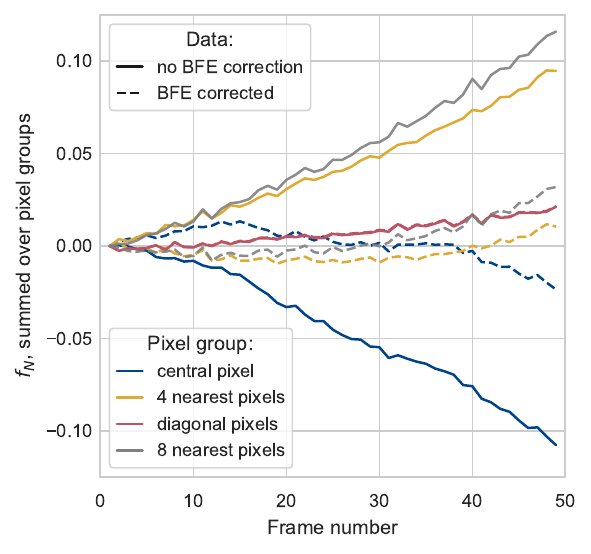}
\caption{Comparison of the median relative flux deviation, $f_N$, in $3\times3$ pixel regions around pixel-centered spots before and after attempting to correct for BFE. Solid lines indicate CNL-corrected data (same as Figure \ref{fig:flux deviation}), dashed lines indicate the additional BFE-corrected data, and color indicates the pixels being summed. Note that the BFE correction has very little impact on the flux in the diagonal pixels, so the dashed and solid pink lines are nearly indistinguishable on this scale. \label{fig:flux deviation corrected}}
\end{figure}

\begin{figure}[t!]
\centering
\includegraphics[width=\columnwidth]{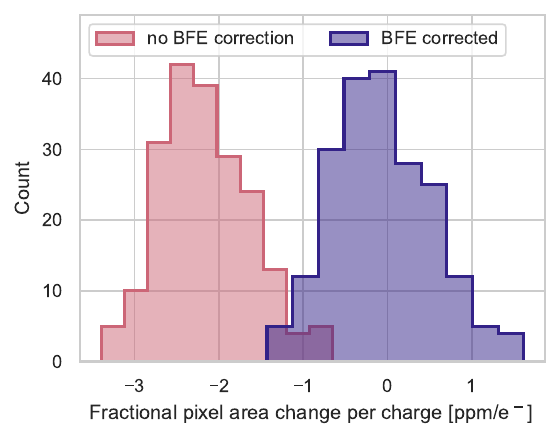}
\caption{Best-fit $B$ coefficients, representing the fractional pixel area change per charge contrast, for the pixel-centered spots before (pink) and after (indigo) applying the BFE correction. The mean values of $B$ are $-2.18\pm0.04$ ppm/e$^-$\ for the uncorrected data and $-0.08\pm0.04$ ppm/e$^-$ for the BFE-corrected data.
\label{fig:B hist compare}}
\end{figure}

Figure~\ref{fig:flux deviation corrected} shows the relative flux deviation around pixel-centered spots for the BFE-corrected ramps in comparison to the uncorrected ramps. 
The correction results in slopes closer to zero for the center pixel and the four nearest neighboring pixels, but has little impact on the diagonal pixels. We quantify the improvement by comparing the best-fit $B$, representing the change in pixel area due to charge contrast. The $B$ fits for both samples are shown Figure~\ref{fig:B hist compare}. The mean values of $B$ are $-0.08\pm0.04$ ppm/e$^-$ for the BFE-corrected data and $-2.18\pm0.04$ ppm/e$^-$\ for the uncorrected data, which is about a factor of 25 improvement. However, Figure~\ref{fig:flux deviation corrected} indicates that higher-order nonlinearity in the uncorrected data remains after BFE correction. The flux is slightly over-corrected for early frames and under-corrected for later frames. 

\begin{figure}[t!]
\centering
\includegraphics[width=\columnwidth]{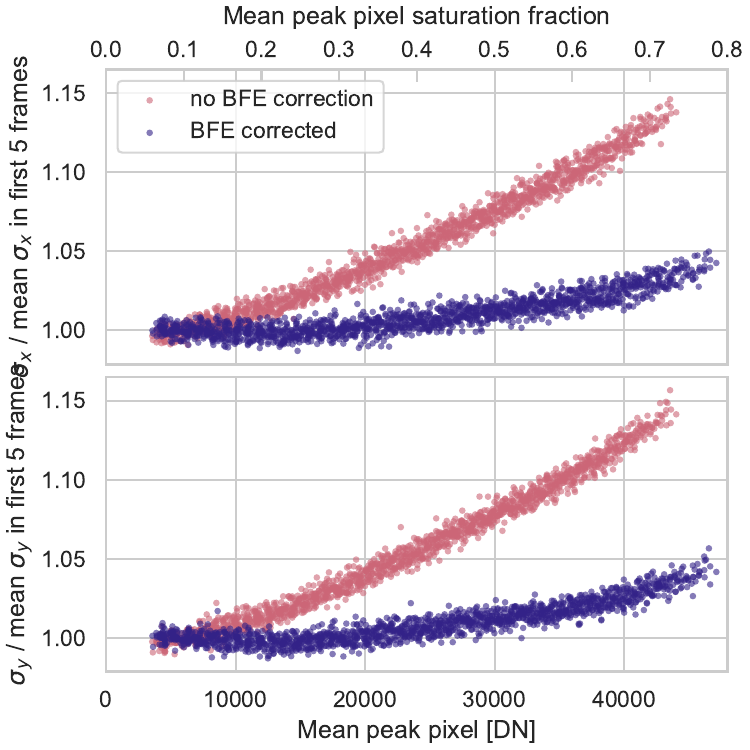}
\caption{Comparison before (pink) and after (indigo) BFE correction of the relative size of the ePSF per frame in the detector $x$ and $y$ pixel directions vs.\ the mean maximum pixel value of spots sampled to fit the ePSF.  \label{fig:epsf size compare corrected}}
\end{figure}

In Figure~\ref{fig:epsf size compare corrected}, we compare the ePSF sizes as a function of signal between the BFE-corrected and uncorrected data. The change in size is substantially reduced after BFE correction, but as with the flux deviation result, we find evidence that the BFE is nonlinear with charge. There is a slight negative trend in the BFE-corrected spot sizes at low signals, corresponding to the over-correction in early frames. At higher signal, the spot sizes increase by $\sim3-5\%$ due to the under-correction. 

\section{Discussion} \label{sec:discussion}

We have presented multiple measurements of the BFE in a spare Roman H4RG detector (SCA \scanum). First, we analyzed the detector flat field statistics following the procedure of \citet{Hirata2020} and \citet{Freudenburg2020} to derive an IPNL matrix describing the coupling between pixels due to accumulated charge, which is dominated by the BFE. Next, we used spot images on the same detector to characterize the impact of BFE around high contrast pixels. We demonstrated that the residual nonlinearity after correcting for classical nonlinearity is consistent with BFE expectations; charge deflection away from pixels with previously accumulated charge results in bright pixels showing decreasing flux of charge over the course of an exposure, which is conserved by increasing flux to the neighboring, lower signal pixels. We also showed that this charge migration corresponds to increasing measured PSF size during a single exposure, meaning that the effective PSF for an uncorrected Roman detector is a function of total accumulated signal and contrast between neighboring pixels. Finally, we presented a BFE correction adapted from \citet{Coulton2018} for H4RG detectors. The method corrects most of the effect, but there is apparent remaining nonlinearity that corresponds to a few percent change in spot image size. 

\subsection{BFE metrics}

As described by \citet{PlazasMalagon2024}, the relative flux deviation around bright pixels is related to the fractional change in pixel area per \electron\ of charge contrast, $B$. They also detailed a procedure to translate the IPNL coefficients from flat field statistics to a predicted $B$ assuming specific spot geometry, and found $\sim$18\% disagreement when comparing their spot-based and flats-based $B$ coefficients. For SCA \scanum, we found a mean $B$ coefficient of $-2.18\pm0.04$ ppm/\electron\ using the spot-based relative flux deviation, which we note is consistent with the corresponding central pixel IPNL measurement from flats of $-2.142 \pm 0.007$ ppm/e$^-$. One can also use results of the correction presented above to qualitatively assess the consistency between spots-based and flats-based BFE since the flat field statistics determine the kernel used in the correction of spot images. The mean $B$ coefficient fit to the corrected ramps is reduced to $-0.08\pm0.04$ ppm/\electron, suggesting that flat field IPNL results are generally consistent with the effect observed in the higher contrast spots. 

We have implemented a scalar kernel correction to the BFE, which assumes that the effect is constant with signal and translationally invariant. While the correction removes most of the observed BFE, there is remaining nonlinearity which may be due to higher order BFE. The correction residuals are indicative of the scalar kernel model assumptions breaking down as signal and contrast between pixels increases. Other sources of IPNL (e.g.\ the vertical trailing pixel effect) are not included in this model and may also be affecting the residual change in spot shape. Errors in the classical nonlinearity correction may also appear as a contribution to the relative flux deviation per pixel. Further, H4RG detectors show other non-ideal effects, such as burn-in \citep{Mosby2020}, which are not as well understood and would contribute to residual nonlinearity. Future work should investigate the contributions of these nonlinear effects to the observed BFE and the results of a BFE correction algorithm. 

\subsection{Comparison to expected Roman survey data} \label{sec:survey data comparison}

There are several differences between the TI spot data and what is expected for real future Roman data which must be noted when interpreting our results. Primarily, we cannot obtain a similar spot data set for any of the 18 detectors integrated into Roman's focal plane. The total IPNL has been shown to vary between detectors \citep{Freudenburg2020}, so one should not expect the results for SCA \scanum\ to be exactly the same as any flight detector. There is a relative paucity of long-exposure flats for this detector in the TI, and therefore the CNL correction implemented here is most likely not as robust as what is expected for flight data. Finally, we have analyzed single frame CDS images to characterize BFE, whereas Roman flight data will be grouped into resultant frames and only these resultant frames will be down-linked. Thus, each resultant will exhibit an average BFE over a variable number of frames. Despite these differences between TI and flight data, the analyses presented here may be a framework for characterizing the BFE from stellar images during flight. 

To put these results in the context of stars in Roman's HLWAS, we estimate the magnitudes of stars which will produce comparable peak pixel count rates to the spots in our data. We used \textsc{pandeia} \citep{Pontoppidan2016}, the simulation tool underlying Roman's exposure time calculator, to simulate stars of between 17.5-18.5 mag in the F106, F129, and F158 filters. For each image, we simulate only a star located at the center of the image, using a flat spectrum normalized to the given magnitude in each respective filter. The exposure parameters were set to match the HLWAS medium tier single-exposure images (107 second integration time over 7 unequally spaced resultants; \citealt{ROTAC2025}). We use the resulting count rate measured in \electron/s of the brightest pixel in each image to compare to the count rates of the spot images. For roughly pixel-centered spots, the median count rate in the brightest pixel was $\sim590$ \electron/s, which corresponds to the peak pixel rate for magnitudes of 18.2, 18.0, and 17.8 in F106, F129, and F158, respectively. We note that the TI spots do not replicate the shape of Roman's optical PSF, so while these are comparable peak pixel rates, the contrast with surrounding pixels will be different for real Roman data.

For weak lensing cosmology, one needs to consider how PSF shape errors propagate to galaxy shape errors, since the scale of BFE will be different for relatively bright stars underlying the PSF model compared to fainter galaxies. In Appendix \ref{app:shape bias}, we present estimates of the galaxy shape bias introduced by the BFE following the formalism of \citet{Paulin-Henriksson2008}. The results, shown in Figure \ref{fig:shape bias}, scale with the ratio of the squared PSF radius to galaxy radius, $\left( R_{PSF} / R_{gal} \right)^2$. For a ratio of $R_{PSF} / R_{gal}\sim 1/2$, the estimated galaxy ellipticity multiplicative bias exceeds $\sim1\%$ for signal levels $>20\%$ of full well based on the uncorrected ramps. This corresponds to approximately the total signal accumulated in the brightest pixel for a stellar magnitude of $\sim$19 in the medium and wide tiers of the HLWAS with 107 s exposures, and a magnitude of $\sim20$ in the deep tier with 294 s exposures. Meanwhile, at this signal level the BFE-corrected data results in a multiplicative bias of $\lesssim 0.2\%$.

While there are substantial differences between the TI spot images and real Roman stellar images, the estimates in Figure \ref{fig:shape bias} suggest that fitting to stars at high signal will bias PSF shape beyond Roman's requirements, which are to calibrate PSF ellipticity and size to $\leq 5.7\times10^{-4}$ and $\leq 7.2\times10^{-4}$, respectively. Thus, future studies are required to determine the ideal BFE mitigation or correction strategy for Roman survey data. For example, the LSST pipeline utilizes a correction for higher order BFE by more directly modeling the electrostatics in a pixel and computing the resulting pixel boundary shifts \citep{Broughton2024}. Given the residual nonlinearity after the BFE correction tested in this paper, a scalar kernel model of BFE may be insufficient to correct the effect at high signal levels as the assumptions of a signal-independent, translationally invariant kernel break down. Future work should investigate methods to implement a higher order BFE correction for Roman and the implications of various BFE mitigation strategies for resulting cosmology.

\begin{acknowledgements}
We thank Christina Hedges, Andrés Plazas, and Alex Broughton for helpful discussions.
This work was supported by NASA under award number 80GSFC21M0002. 
\end{acknowledgements}

\section*{Data Availability}
A public release of TI data is planned at the time of writing, but the date is to be determined. In the meantime, the data used in this work may be available upon reasonable request. Inquiries about TI data should be directed to Ami Choi and Joshua Schlieder.\footnote{ami.choi@nasa.gov, joshua.e.schlieder@nasa.gov}


\vspace{5mm}
\software{astropy \citep{AstropyCollaboration2022}, 
          photutils \citep{Bradley2025},
          numpy \citep{Harris2020},
          matplotlib \citep{Hunter2007}
          }



\appendix

\section{Impact of the CNL correction} \label{app:cnl}

\begin{figure}[t!]
\centering
\includegraphics[width=0.5\columnwidth]{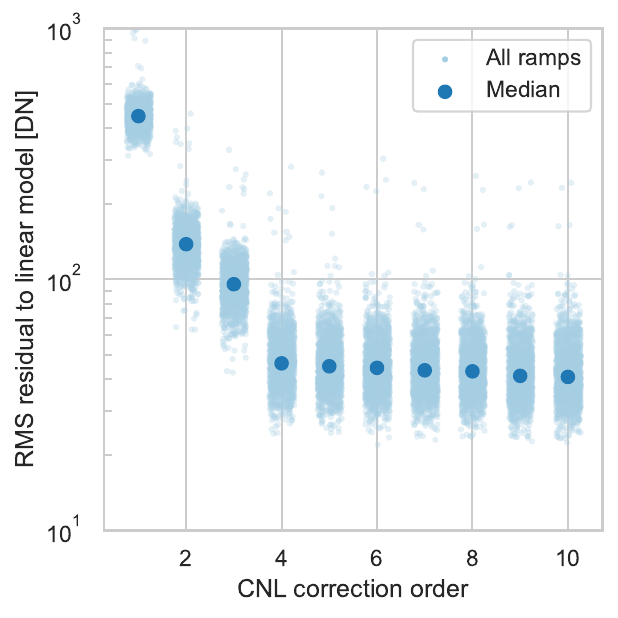}
\caption{Root mean square residuals of CNL-corrected diffuse exposure ramps to a linear fit for each CNL polynomial order tested. Light blue points are the residual for each pixel tested in a $32\times32$ pixel region at the center of the SCA \scanum\ (with random horizontal offsets for display purposes only), and the darker blue points are the median of all pixels per correction order. The residuals decrease by about a factor of 10 as one increases the correction order from 1 to 4, then remains stable to within a few DN over increasing correction order.
\label{fig:cnl order residuals}}
\end{figure}

In the work presented above, we have assumed that the nonlinearity observed in the pixel ramps after correcting for CNL is predominantly caused by BFE. Thus, the interpretation of the results is dependent on the accuracy of the CNL solution, which will vary based on the polynomial order chosen for the correction. To determine an appropriate polynomial order for the CNL correction, we fit corrections to a sample of 1024 pixels from 1st to 10th order, using the two exposures of diffuse illumination from the TI and the same methodology described above in Section \ref{sec:cnl} \citep{Brandt2025}. For each pixel and order, we calculated CNL-corrected ramps and fit a straight line using least squares minimization. Figure \ref{fig:cnl order residuals} shows the root mean square residuals of the corrected ramps to the linear model as a function of polynomial order. The residuals decrease by about an order of magnitude as correction order increases from 1 to 4, and are then relatively stable up to order 10. We therefore chose a 4th order polynomial correction for our main results to avoid over-fitting to noise in the TI exposures.

\begin{figure*}[t!]
\centering
\includegraphics[width=\textwidth]{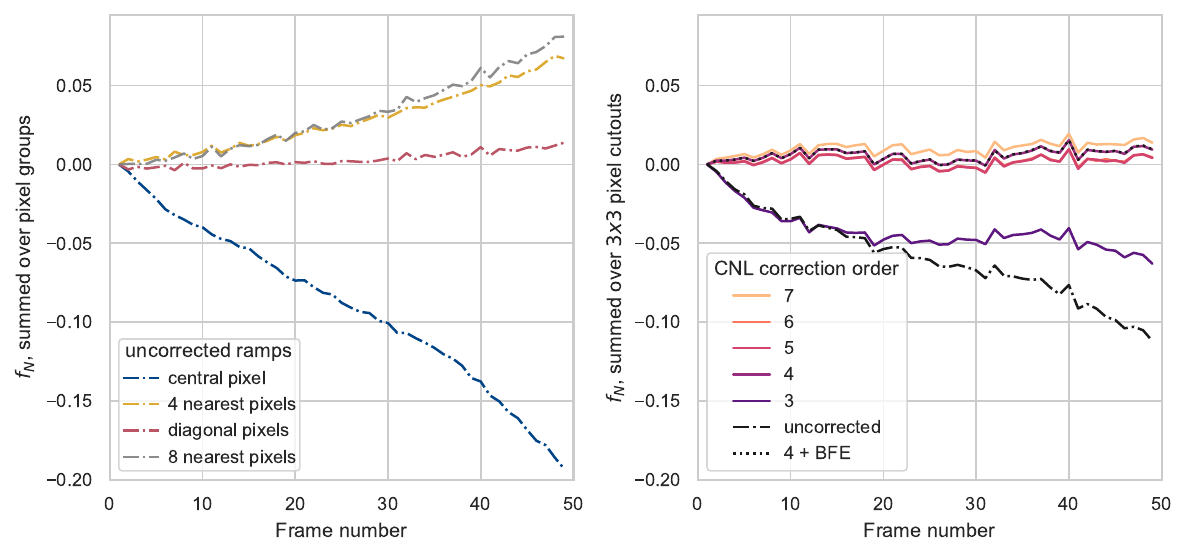}
\caption{Left: the median relative flux deviation, $f_N$, in $3\times3$ pixel regions around pixel-centered spots in the data with no CNL correction applied, summed over pixel groupings in the same fashion as Figure~\ref{fig:flux deviation corrected}. 
Right: the sum of $f_N$ over the full 9 pixel cutouts for the data with no correction compared to the CNL correction with order ranging from 3-7. 
Without correcting for CNL, the apparent flux-loss in the central pixel is not conserved by increasing flux in the surrounding 8 pixels. For CNL corrections with polynomial orders $\geq4$, the total flux is roughly conserved and variation in the total flux per order is comparable to the noise from frame to frame. The 4th order CNL + BFE correction is indicated by the dotted black line, showing that the BFE correction does not alter the total $f_N$ on the scale of this plot.
\label{fig:cnl fn compare}}
\end{figure*}

To check how the CNL correction step has affected the BFE results, we calculated the relative flux deviations around pixel-centered spots using the dark-subtracted spot ramps prior to CNL corrections and with several different choices of polynomial order for the CNL. The results are shown in Figure~\ref{fig:cnl fn compare}. 
If the total spot flux is constant over time and the charge migration due to BFE is limited to $\sim$1 pixel radius, then the sum of $f_N$ over the $3\times3$ pixel cutouts should be zero over all frames. 
As shown in Figure~\ref{fig:cnl fn compare}, the apparent signal loss over time in the central pixel is not fully conserved by increased signal in the surrounding pixels for both the uncorrected data and the 3rd order polynomial CNL correction. However, for CNL correction orders $\geq4$, the sum is close to zero, and it is unchanged by the additional BFE correction. There is some variation in total $f_N$ depending on CNL order, but the scale of the variation is similar to the noise in the $f_N$ measurement per frame. 
We interpret this as an indication that the CNL correction is functioning as expected: the correction removes the component of nonlinearity that is observable under flat illumination, and the residual nonlinearity is consistent with total signal-conserving inter-pixel effects. 
Observed deviations from zero total $f_N$ in this data may be a consequence of imperfect CNL correction, or there may be other underlying trends due to unconstrained laser fluctuations and detector effects such burn-in.

\section{Shape Measurement Bias Estimation} \label{app:shape bias}

In the context of weak lensing cosmology, the goal of a BFE calibration and correction would be to mitigate the bias propagated from PSF errors to galaxy shape errors and subsequent gravitational lensing shear measurements. It is therefore useful to consider how the ePSF growth seen in this work would correspond to systematic errors in shape measurements. To do this, we parameterize shapes by ellipticity $\epsilon$ and radius parameter $R$. For 2D elliptical Gaussian fits, the ellipticity and squared radius are related to the standard deviation along the major and minor axes as
\begin{equation}
    \epsilon = \frac{\sigma^2_{major}-\sigma^2_{minor}}{\sigma^2_{major}+\sigma^2_{minor}},\ 
    R^2 = \sigma^2_{major}+\sigma^2_{minor}.
\end{equation}
Following Eq.\ 13 of \citet{Paulin-Henriksson2008}, we assume that the systematic error in galaxy ellipticity measurement is related to PSF shape errors by
\begin{equation}
    \delta \epsilon^{sys} \simeq \left(\epsilon_{gal} - \epsilon_{PSF}\right) \frac{\delta R_{PSF}^2}{R_{gal}^2} - \left(\frac{R_{PSF}}{R_{gal}}\right)^2 \delta \epsilon_{PSF} ,
\end{equation}
where $\epsilon_{gal}$ and $R_{gal}$ are the true galaxy shape parameters, $\epsilon_{PSF}$ and $R_{PSF}$ are the PSF shape parameters, and $\delta \epsilon_{PSF}$ and $\delta R_{PSF}^2$ are the error in PSF shape parameters. Thus, the galaxy ellipticity measurement errors can be categorized by a multiplicative bias, 
\begin{equation}
    m \sim \frac{\delta R_{PSF}^2}{R_{gal}^2} , 
    \end{equation}
and an additive bias, 
\begin{equation}
    c \sim - \left(\frac{R_{PSF}}{R_{gal}}\right)^2 \delta \epsilon_{PSF}. 
\end{equation}

\begin{figure}[t!]
\centering
\includegraphics[width=0.5\columnwidth]{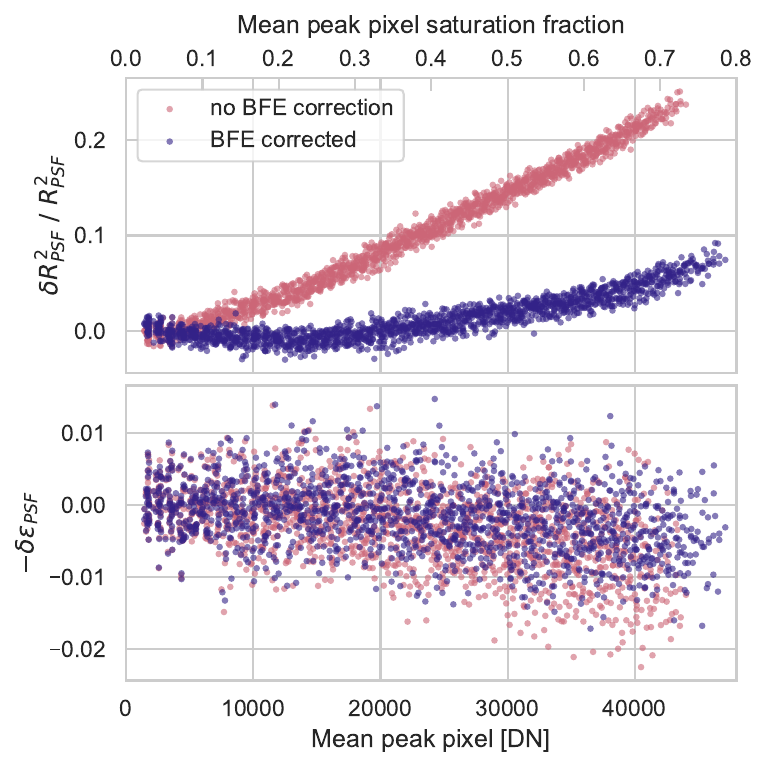}
\caption{Estimates of shape bias imparted by changing ePSF shape as a function of signal for the CNL corrected (pink) and CNL+BFE corrected (indigo) data. The top panel is proportional to the multiplicative bias, $m$, and the bottom panel is proportional to the additive bias, $c$, both by a factor of $(R_{PSF}/R_{gal})^2$.}
\label{fig:shape bias}
\end{figure}

For the TI spot data, we estimate these PSF shape errors from the difference in ePSF ellipticity and size over the course of the ramps compared to early frames, which we assume are the closest to the true optical spot profiles. Figure \ref{fig:shape bias} shows $\delta R_{PSF}^2/R_{gal}^2$ and $-\delta \epsilon_{PSF}$ --- both of which are related to the multiplicative and additive bias parameters, respectively, by a factor of $(R_{PSF}/R_{gal})^2$ --- for the CNL corrected ramps and CNL+BFE corrected ramps. The multiplicative bias follows the same trend as the ePSF size measurements in Figure \ref{fig:epsf size compare corrected} since it is simply related to the width of the 2D Gaussian fits. To relate these results to expected bias on galaxy shapes, one would need to choose a ratio of PSF size to galaxy size and multiply by $(R_{PSF}/R_{gal})^2$. Thus for any galaxy larger than the PSF, the fractional bias will be smaller than the plotted values. However, we note that due to differences between the TI experiment and real Roman survey data (discussed in Section \ref{sec:survey data comparison}), these should not be interpreted as precise predictions of the galaxy shape bias, and the real bias will need to be investigated with on-sky PSF measurements. 

\bibliography{references}{}
\bibliographystyle{aasjournal}



\end{document}